# Primary and secondary scintillation measurements in a xenon Gas Proportional Scintillation Counter

L.M.P. Fernandes[1], E.D.C. Freitas[1], M. Ball[2], J.J. Gómez-Cadenas[2], C.M.B. Monteiro[1], N. Yahlali[2], D. Nygren[3] and J.M.F. dos Santos[1]

[1] *Instrumentation Centre, Physics Department, University of Coimbra, P-3004-516 Coimbra, Portugal*
[2] *Instituto de Física Corpuscular, E-46071 Valencia, Spain*
[3] *Lawrence Berkeley National Laboratory, Berkeley, CA 94720, USA*

*E-mail: pancho@gian.fis.uc.pt*

**ABSTRACT**

NEXT is a new experiment to search for neutrinoless double beta decay using a 100 kg radio-pure high-pressure gaseous xenon TPC. The detector requires excellent energy resolution, which can be achieved in a Xe TPC with electroluminescence readout. Hamamatsu R8520-06SEL photomultipliers are good candidates for the scintillation readout. The performance of this photomultiplier, used as VUV photosensor in a gas proportional scintillation counter, was investigated. Initial results for the detection of primary and secondary scintillation produced as a result of the interaction of 5.9 keV X-rays in gaseous xenon, at room temperature and at pressures up to 3 bar, are presented. An energy resolution of 8.0% was obtained for secondary scintillation produced by 5.9 keV X-rays. No significant variation of the primary scintillation was observed for different pressures (1, 2 and 3 bar) and for electric fields up to 0.8 V $cm^{-1}$ $torr^{-1}$ in the drift region, demonstrating negligible recombination luminescence. A primary scintillation yield of 81 ± 7 photons was obtained for 5.9 keV X-rays, corresponding to a mean energy of 72 ± 6 eV to produce a primary scintillation photon in xenon.

## 1. INTRODUCTION

NEXT stands for Neutrino Experiment with a Xenon TPC, a search for neutrinoless double-beta decay (0νββ) [1]. The experiment will use the $^{136}$Xe isotope in a 100 kg high pressure (~10 bar) TPC (time projection chamber), to be operated at room temperature in the Canfranc Underground Laboratory. The unambiguous observation of 0νββ would demonstrate that the neutrino has a Majorana nature. This would represent a breakthrough of new physics, beyond the Standard Model. The TPC should combine an excellent energy resolution with a unique topological signature to achieve high sensitivity to a light Majorana neutrino.

Xenon offers several advantages for 0νββ experiments. As a noble gas, it can be used for tracking particles. It does not have long-lived radioactive isotopes other than $^{136}$Xe, which decays by double-beta. This isotope has a relatively high abundance in natural xenon (8.9%) and can be easily enriched by centrifugation at a reasonable cost.

The $^{136}Xe \rightarrow {}^{136}Ba$ transition has a high Q-value, 2457.83(37) keV [2], allowing a reduction of the background resulting from lower energetic gamma-rays emitted by other radioactive materials by means of pattern recognition of the events in the offline analysis. Furthermore, xenon can be used at the same time as the detection medium for the charged particles released in the decay. Another interesting property of pure Xe is the large amount of both primary ionization and primary VUV scintillation (~175 nm) induced by the radiation interaction.

The detector concept requires excellent energy resolution (< 1% at 2.46 MeV). This is essential not only to reduce the tail of the double-beta decay with neutrino emission ($2\nu\beta\beta$) spectrum from overlapping the region of interest of the $0\nu\beta\beta$ spectrum, but also to prevent the contamination of the region of interest by the most severe gamma-ray background (2614 keV from $^{208}Tl$ and 2447 keV from $^{214}Bi$). Furthermore, external backgrounds should also be reduced using topological properties of the $0\nu\beta\beta$ events.

For half a century it has been known that secondary scintillation, also called electroluminescence, provides a mechanism for high gain with very low fluctuations [3]. This technique offers large signals with negligible electronic noise, and is the optimum amplification technique for this kind of experiment. Therefore, it is of great importance, especially in experiments with very low event rates and/or high background levels such as $0\nu\beta\beta$ experiments, to use the secondary scintillation signal rather than the signal from either unamplified primary ionization or avalanche ionization [4]. This is the technique to be used in NEXT, with a nominal xenon pressure of 10 bar [1].

The proposed detector design for NEXT, called Separated Optimized Function TPC (SOFT) approach, is based on a specific readout that separates the technologies for pattern recognition and energy measurement [4]. Electroluminescence photons emitted towards the hemisphere of the anode can be used to recognise the specific track pattern of the two electrons emitted in the double beta decay. The technology does not require excellent energy resolution capabilities but does require robust pattern recognition and the capability to separate nearby hits. The photons emitted in the opposite direction can be detected with a series of PMTs mounted behind the cathode. Here, energy resolution and the identification of the start-of-event signal (t0) are the major needs, the latter being determined by the primary scintillation. Due to a uniform light distribution at the cathode, the whole area does not have to be covered. However the coverage has to guarantee a good identification of the t0 signal to ensure a full three-dimensional event reconstruction. Therefore the precise knowledge of the expected primary and secondary light densities is crucial to optimise the technology for these two tasks.

The PMTs considered for energy readout are from Hamamatsu R8520-06SEL series [5]. A similar type of PMT, R8520-06-AL, was developed for the double phase detector of the XENON collaboration and optimized for cryogenic operation [6,7]. This type of PMT, which is square shaped with a bialkali (Rb-Cs-Sb) photocathode and a quartz window, presents a quantum efficiency of about 30% at 175 nm. The PMTs are compact (1 $in^2$ active area, 3.5 cm long), have 10 multiplication stages (dynodes) and reach a maximum gain of a few $10^6$. The PMT investigated has a gain of $1.7\times10^6$ for a PMT bias of 800V, according to the manufacturer datasheet. The PMT, operating at room temperature, is able to detect a small number of UV photons.

The study of the performance of such PMTs for the detection of primary and secondary scintillation produced in xenon at room temperature is an important part of the NEXT program. For this purpose, we built a xenon Gas Proportional Scintillation Counter (GPSC) [8] equipped with a R8520-06SEL PMT as VUV photosensor. In such a detector, primary electrons released by ionization of the gas medium drift under an external electric field, below the Xe scintillation threshold, towards a region between two parallel meshes separated by a few mm. In this region, the so-called secondary scintillation region, the electric field is such that the electron energy is kept below the Xe ionization threshold but high enough to excite Xe atoms. The de-excitation of Xe results in isotropic emission of secondary scintillation photons of 175 ± 10 nm, which are detected by a photosensor. This multiplication process presents a linear dependence on the applied electric field [9,10] and smaller statistical fluctuations, resulting in improved energy resolution when compared to charge avalanche processes [8].

In this work, we report the results obtained with such a GPSC for 5.9 keV X-rays absorbed in the xenon. The results for the detection of both primary and secondary scintillation are presented and compared with other high performance GPSCs equipped with standard PMTs.

## 2. EXPERIMENTAL SETUP

The GPSC investigated is schematically depicted in Fig. 1. The R8520-06SEL PMT was glued with low vapour pressure epoxy (TRA-CON 2116) to the pressure vessel on the anode plane. The GPSC has an aluminized Kapton window, a 3 cm thick drift region between the window and mesh G1, and a scintillation gap of 0.5 cm between mesh G1 and mesh G2, which covers the PMT window, used as the anode plane. A radioactive source is positioned outside the chamber, on top of the detector window. The radiation window and mesh G1 are biased to negative high voltage, $-HV_0$ and $-HV_1$, while mesh G2 and the detector body are connected to ground. The PMT is operated with positive high voltage. A Macor piece is used to hold and provide electric insulation to the radiation window and mesh G1. Vacuum sealing is achieved by means of low vapour pressure epoxy. The drift electric field is determined by $HV_1$-$HV_0$ and the scintillation electric field by $HV_1$.

The GPSC was pumped to vacuum pressures of about $10^{-6}$ mbar prior to xenon (99.999% pure) filling. The GPSC was operated at room temperature, with the gas circulating by convection through SAES Getters St707 operated at 180ºC.

The study includes gain and energy resolution measurements for secondary scintillation as a function of the reduced electric fields (E/p) in the drift and scintillation regions, and also primary scintillation amplitude measurements as a function of the drift field. The measurements were made with both a digital oscilloscope and a multichannel analyser (MCA) and were performed at gas pressures of 1, 2 and 3 bar. PMT signals were fed through a low-noise charge sensitive preamplifier (Canberra Model 2005, with a charge conversion gain of 4.5 mV/pC) to a spectroscopy amplifier (Tennelec TC243, with coarse gain selectable between 5 and 2000 and shaping time constants between 0.5 and 12 μs). Then signals were sent to a digital oscilloscope (Tektronix TDS 2022B) or were pulse-height analysed by a 1024-channel MCA (Nucleus PCA II). Measurements were made with shaping time constants sufficiently large to integrate fully over variations in collection time.

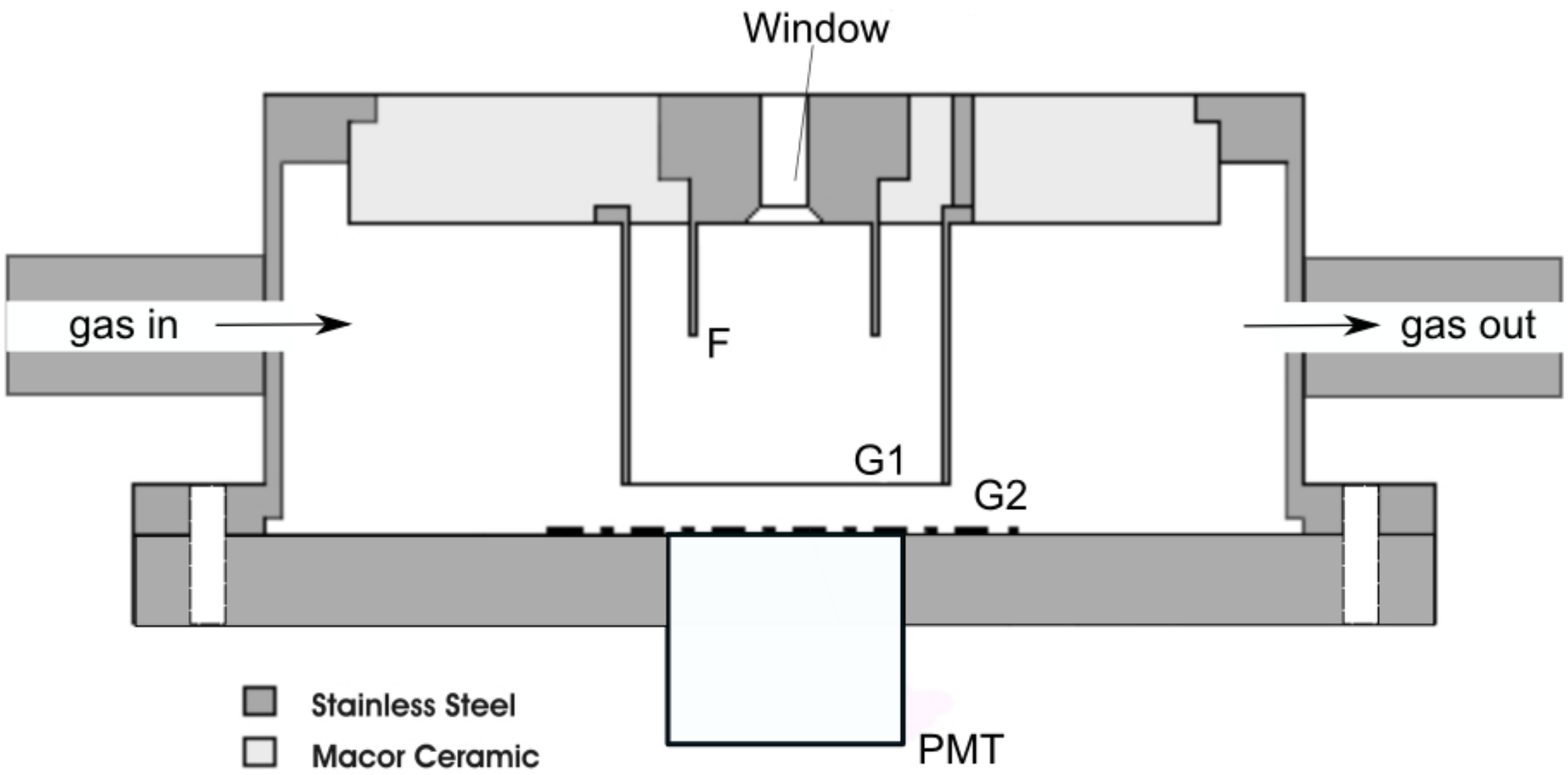

Fig. 1. Schematic of the xenon GPSC with a R8520-06SEL PMT as VUV photosensor.

## 3. ELECTROLUMINESCENCE MEASUREMENTS

The response of the R8520-06SEL PMT to the electroluminescence produced within the xenon GPSC was investigated. The amplitude and energy resolution of the scintillation pulses resulting from the interaction of 5.9 keV X-rays, emitted by a $^{55}$Fe radioactive source, were determined for different electric fields in the drift and scintillation regions of the GPSC and for gas pressures up to 3 bar.

A thin chromium film was placed between the radioactive source and the window to efficiently reduce the interaction of 6.4 keV X-rays (Mn $K_{\alpha}$ line) in the gas volume, while the absorption of 5.9 keV X-rays (Mn $K_{\beta}$ line) in the chromium film is not so significant. A typical pulse-height distribution obtained for 5.9 keV X-rays is depicted in Fig. 2. The distribution was obtained at atmospheric pressure for optimal reduced electric fields in the GPSC drift and scintillation regions of 0.6 and 5.0 V cm$^{-1}$ torr$^{-1}$, respectively, and for a PMT bias voltage of 660 V, which corresponds to a gain of about $3\times10^{5}$ according to the manufacturer datasheet. An energy resolution of 8.0% (FWHM) was obtained for the 5.9 keV X-ray peak, demonstrating high performance, similar to that obtained with GPSCs instrumented with larger PMTs [11,12].

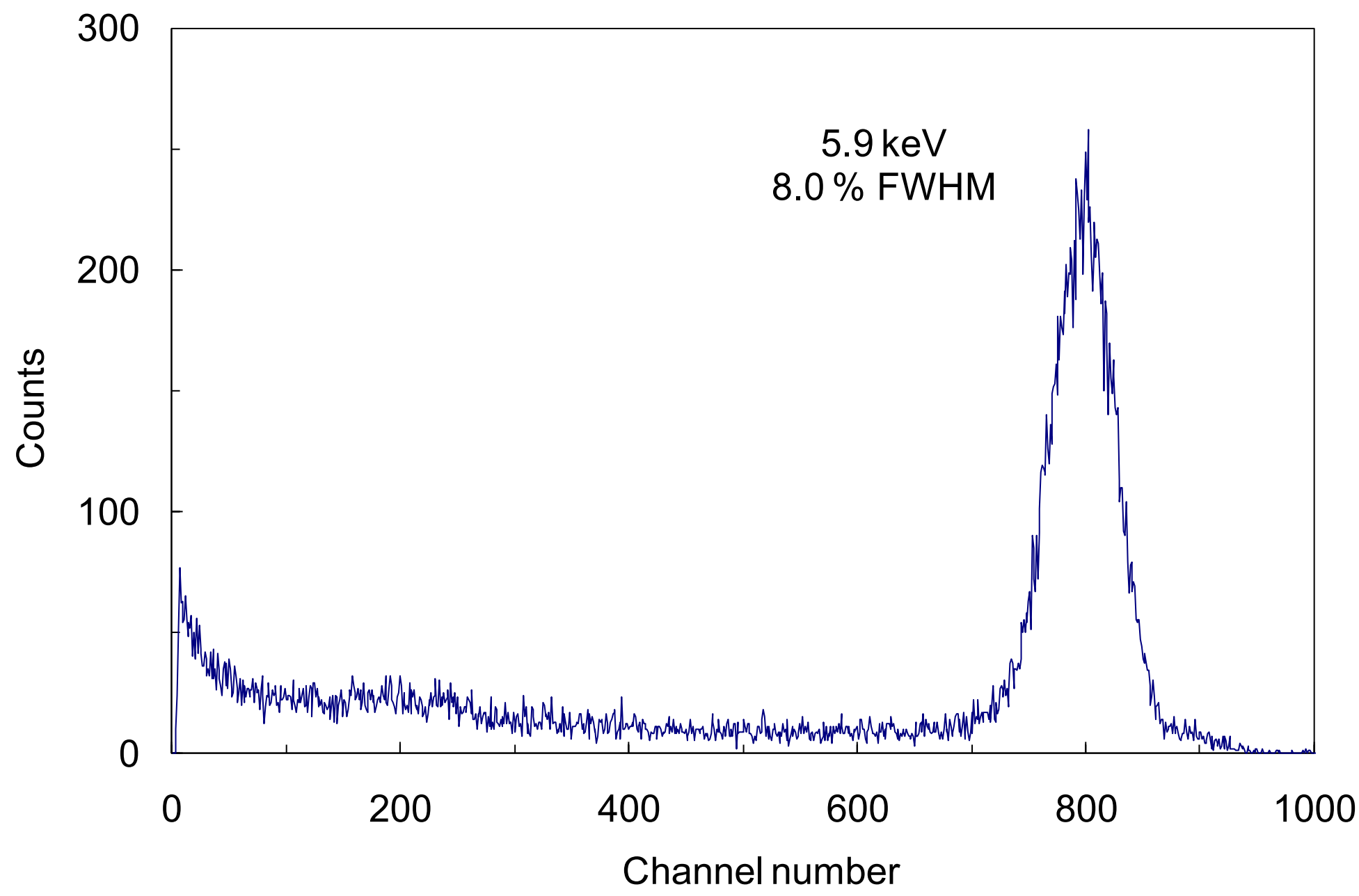

Fig. 2. Pulse-height distribution for 5.9 keV X-rays absorbed in the xenon GPSC. The PMT was biased to 660V. Optimal electric field values of 0.6 and 5.0 V $cm^{-1}$ $torr^{-1}$ were used in the drift and scintillation regions, respectively.

Since the statistical fluctuations associated to the production of VUV scintillation can be neglected, the energy resolution of a conventional GPSC is determined by the statistical fluctuations occurring in the primary ionization processes and in the photosensor. For a PMT photosensor, the energy resolution $R$ (FWHM) is approximately [8]:

$$R = 2.355\sqrt{\frac{F}{N} + \frac{2}{N_e}} \quad (1)$$

where $N$ is the average number of primary electrons produced per incident X-ray photon, $F$ (Fano factor) is the relative variance of $N$, and $N_e$ is the average number of photoelectrons produced in the photosensor per X-ray photon absorbed in the drift region. Taking into account that $N = E_x/w$ ($E_x$ being the X-ray photon energy and $w$ the mean energy to produce a primary electron) and defining the number of photoelectrons produced per primary electron, $L = N_e/N$, the energy resolution can be given by:

$$R = 2.355\sqrt{\frac{w}{E_x}\left(F + \frac{2}{L}\right)} \quad (2)$$

$L$ is a parameter that describes the photosensor performance. For the present PMT, $L$ = 19 assuming $F$ = 0.2 [8]. This value is in good agreement with calculations presented in section 5.

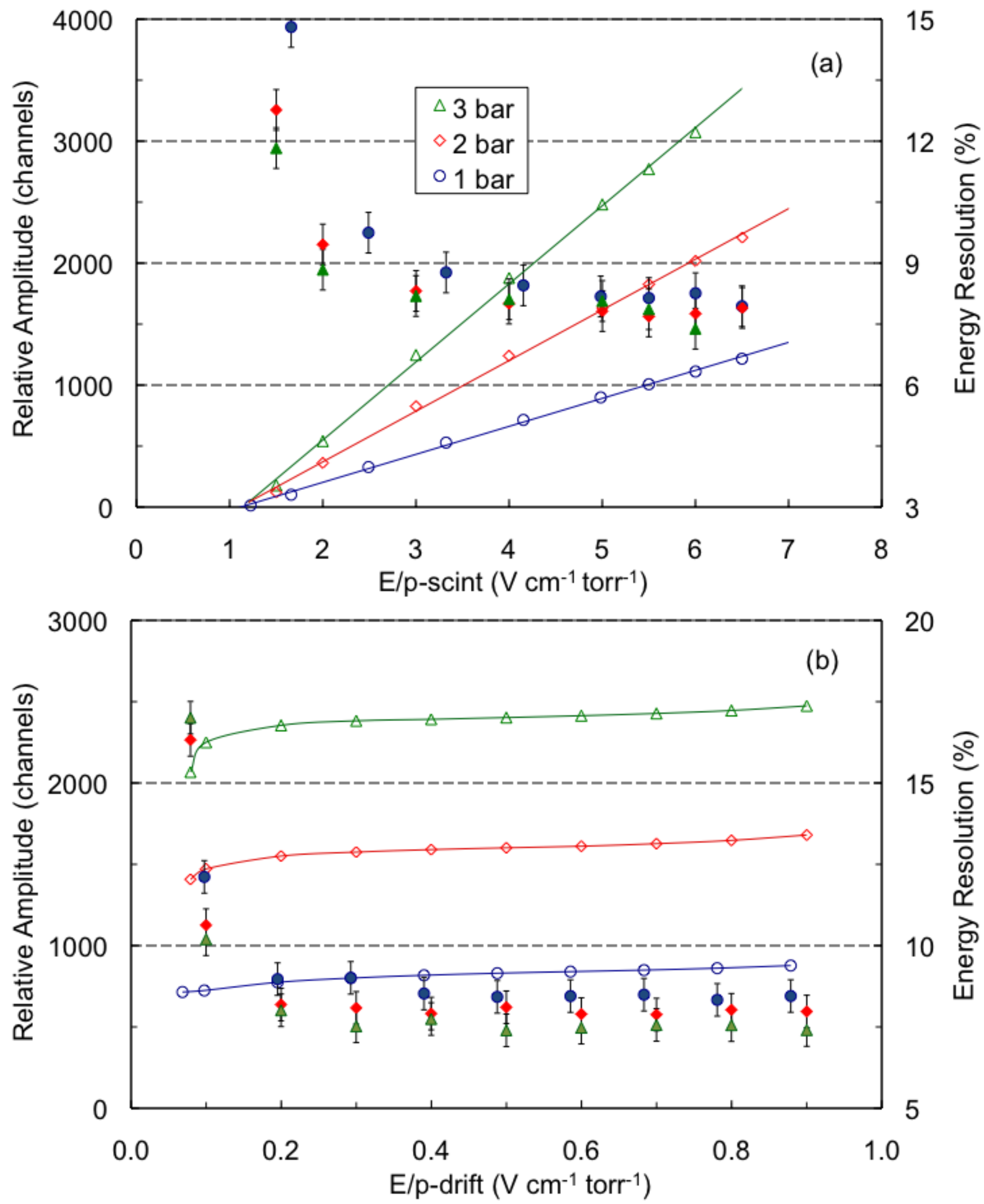

Fig. 3. Pulse amplitude (open symbols) and energy resolution (full symbols) for 5.9 keV X-rays absorbed in the GPSC as a function of: (a) E/p-scint, the reduced electric field in the scintillation region; (b) E/p-drift, the reduced electric field in the drift region. In (a), a fixed drift electric field of 0.5 V cm$^{-1}$ torr$^{-1}$ was used and the PMT was biased to 690 V. In (b), a scintillation electric field of 4 V cm$^{-1}$ torr$^{-1}$ was used and the PMT was biased to 710 V. Results for different gas pressures (1, 2 and 3 bar) are shown.

To obtain this excellent energy resolution, the reduced electric fields in the scintillation region (E/p-scint) and in the drift region (E/p-drift) had to be optimized. The variation of the amplitude and energy resolution with E/p-scint is shown in Fig. 3 (a), while the variation with E/p-drift is shown in Fig. 3 (b). Fig. 3 clearly shows that E/p-scint has to be larger than 4 V cm$^{-1}$ torr$^{-1}$ in order to get the best energy resolution, while E/p-drift has to be larger than 0.2 V cm$^{-1}$ torr$^{-1}$. In both cases, the energy resolution does not have significant variations with pressure. The signal amplitude increases linearly with pressure, as expected. For the same E/p value, the ratio amplitude/pressure is then not significantly dependent on pressure, in accordance with former studies [10].

## 4. PRIMARY SCINTILLATION MEASUREMENTS

Primary scintillation is produced in xenon during the formation of the primary electron cloud following the absorption of radiation and the subsequent thermalisation of the photoelectron and other Auger electrons. The amplitude of primary scintillation pulses is very low and difficult to distinguish from noise. However, by averaging out the noise to a very low level, using a digital oscilloscope, the primary scintillation pulse amplitude can be determined. The oscilloscope is triggered with the secondary scintillation pulse, which takes place a few microseconds later due to the transit time of the primary electrons through the drift region. The amplitude is determined from the average of 128 pulses. Fig. 4 shows typical primary and secondary scintillation pulses, obtained in a Tektronix TDS 2022B oscilloscope. Electric fields of 0.2 and 2.0 V $cm^{-1}$ $torr^{-1}$ were applied to the drift and scintillation regions, respectively. As seen, the primary scintillation pulse is very well distinguished from the noise as a result of the averaging process.

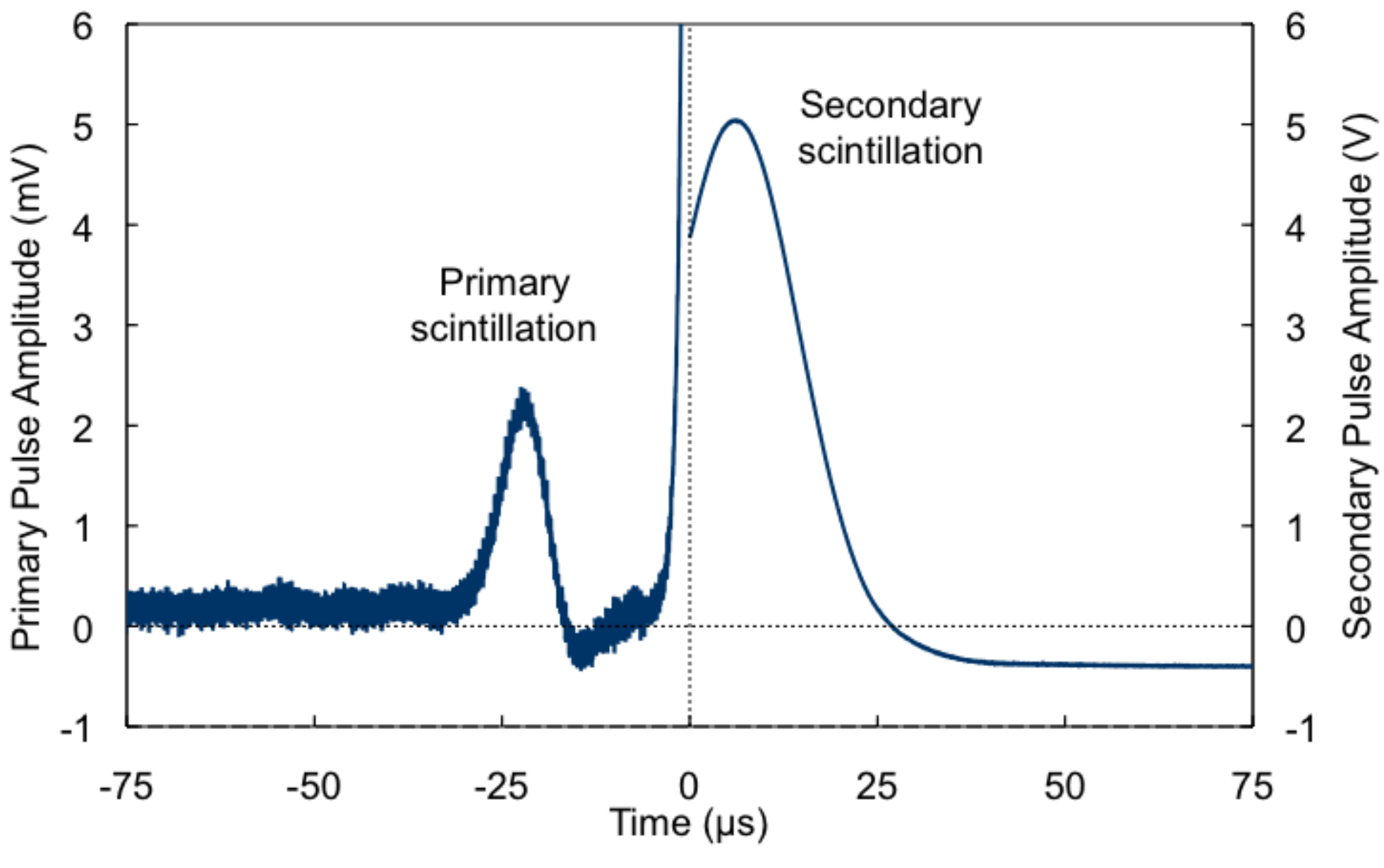

Fig. 4. Typical primary and secondary scintillation pulses observed in the oscilloscope, after averaging 128 pulses, for 5.9 keV X-rays interacting in gaseous xenon. The oscilloscope was triggered at the secondary scintillation pulse.

The primary scintillation pulse amplitude was measured as a function of the drift electric field for pressures of 1, 2 and 3 bar, using an electric field of 2.0 V $cm^{-1}$ $torr^{-1}$ in the scintillation region (Fig. 5). As seen, within the experimental errors, the amplitude variation is not significant for drift electric fields between 0.2 and 0.8 V $cm^{-1}$ $torr^{-1}$ and no significant variation with pressure is observed. Below 0.2 V $cm^{-1}$ $torr^{-1}$ both primary and secondary scintillation pulse amplitudes drop significantly, an effect that has been also observed in Ref. [13]. The drop in the primary scintillation amplitude is not real. This effect can be due to time jitter affecting the averaging in the scope or to the reduction of the electroluminescence pulse amplitude as a result of diffusion and loss of primary electrons to electronegative impurities as the electric field becomes weaker; the subsequent decrease of the signal-to-noise ratio for the trigger pulses adds fluctuations to the signals recorded by the oscilloscope. As the primary signal is within the noise, its averaging is affected. In fact, the amount of the primary scintillation, for low drift fields, could even increase with decreasing electric field due to the presence of additional scintillation resulting from electron-ion recombination. For alpha particles interacting in xenon, the amount of recombination luminescence can reach a fraction above 50% of the total primary scintillation at zero drift electric field [14].

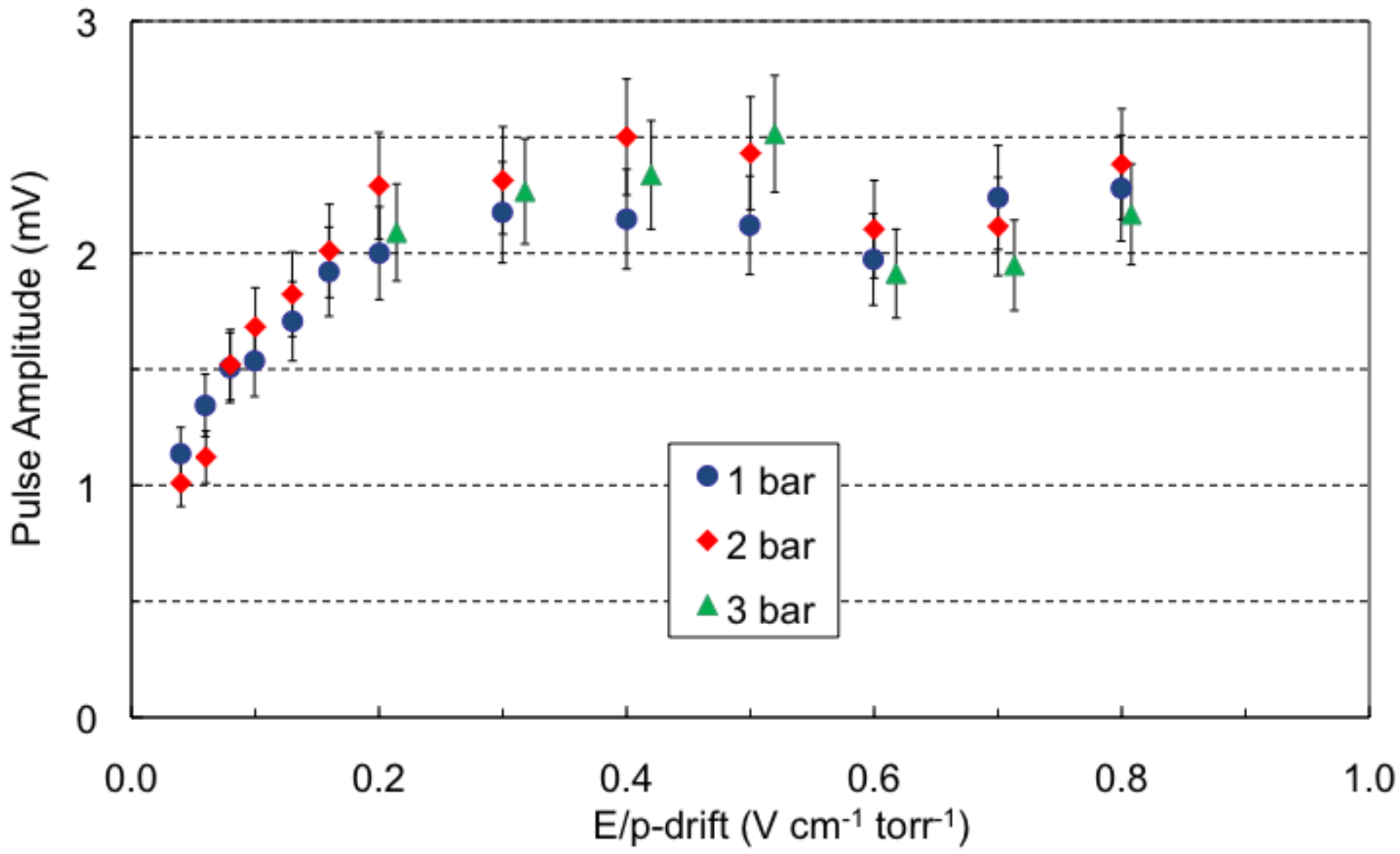

Fig. 5. Primary scintillation pulse amplitude measured in the oscilloscope, for 5.9 keV X-ray interactions in xenon, as a function of the drift electric field.

To overcome the oscilloscope limitation and to look for experimental evidence of recombination luminescence, the electronic settings were optimized in order to detect primary scintillation pulses in the MCA. From the ratio between secondary and primary scintillation pulse amplitudes, we can estimate the region where the pulse-height distribution of primary scintillation should be. To be more sensitive to the detection of the primary scintillation, the amplifier gain and the PMT voltage were increased to higher values.

In order to look for pulse-height distributions of primary scintillation, pulse amplitude distributions were recorded with and without irradiation by 5.9 keV X-rays. This way, the pulse-height distribution for background due to residual visible light entering the chamber could be identified and subtracted (Fig. 6). No electric fields in the GPSC drift and scintillation regions were used. The background rate decreases by improving the light shielding of the detector and by taking data during the night.

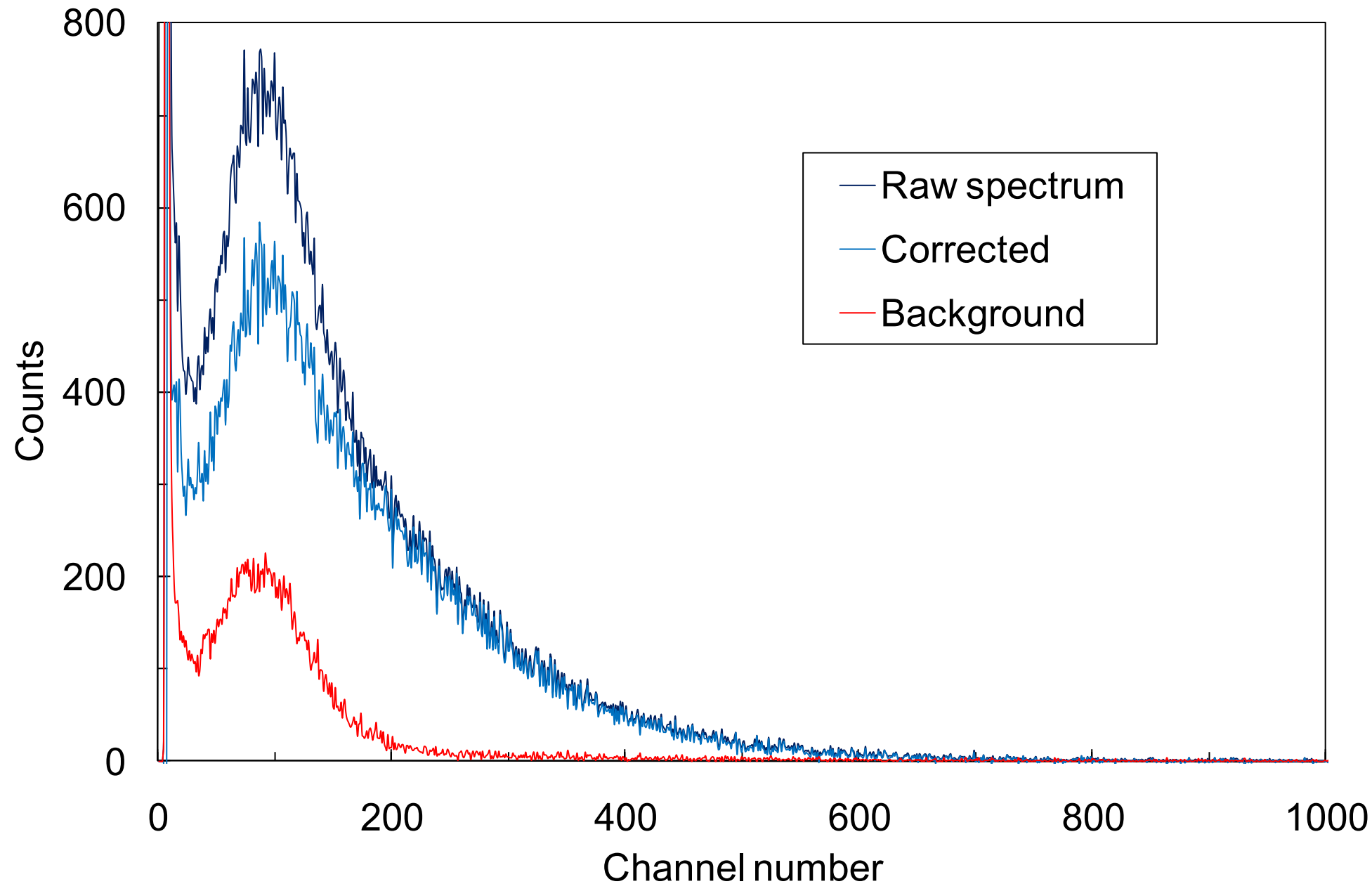

Fig. 6. Pulse-height distribution obtained for 5.9 keV X-rays absorbed in the detector, with no electric fields applied to the drift and scintillation regions and for a PMT bias voltage of 730 V.

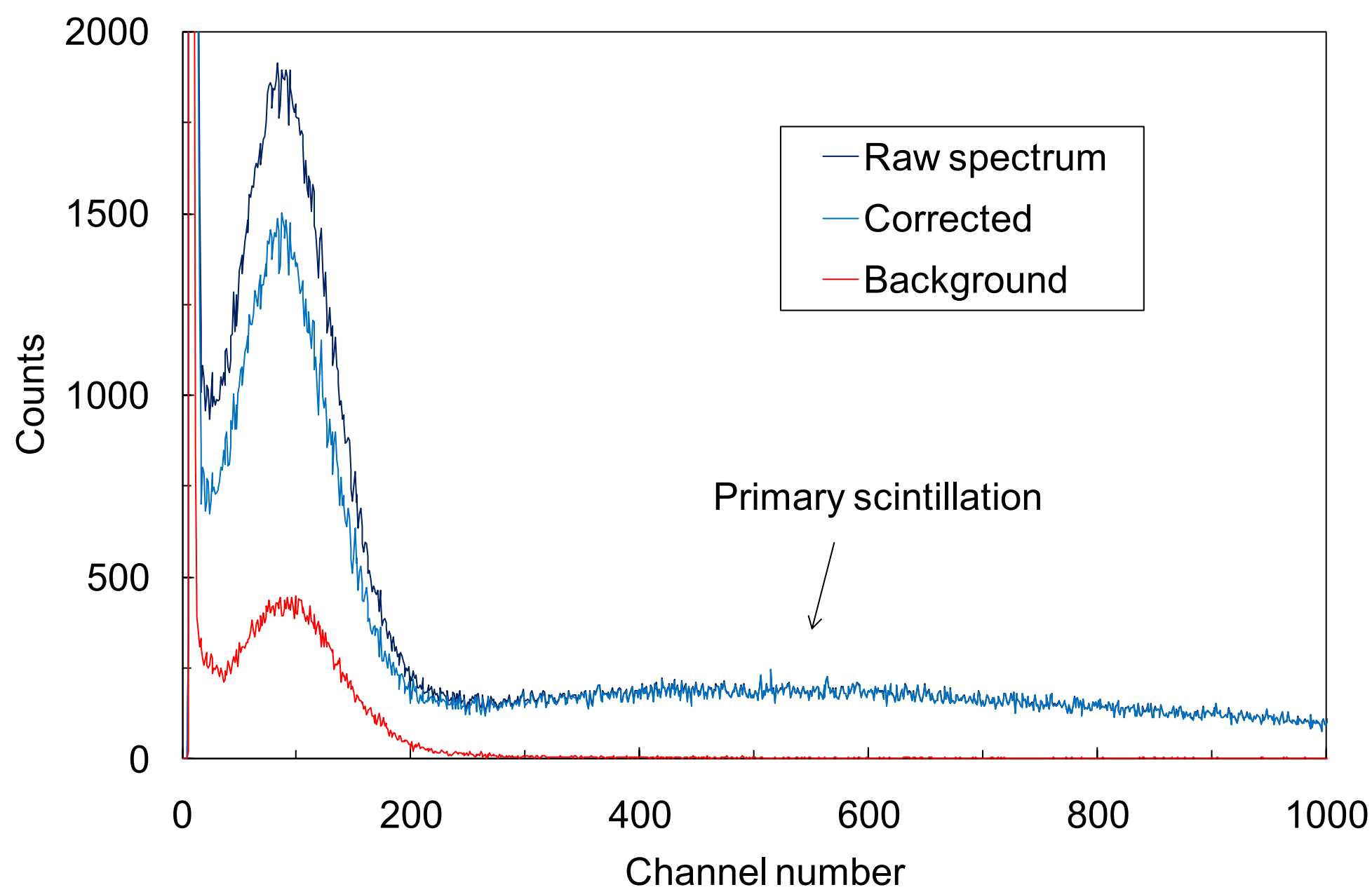

Fig. 7. Pulse-height distribution obtained for the $^{109}$Cd radioactive source, with no electric fields applied to the drift and scintillation regions of the detector and for a PMT bias voltage of 730 V.

The pulse-height distribution obtained after background subtraction has other contributions than primary scintillation due to 5.9 keV X-rays absorbed in xenon. In fact, a similar distribution is obtained in the low-energy region by irradiating the detector with 22.1 and 25.0 keV X-rays from a $^{109}Cd$ radioactive source. The distribution in the higher energy region corresponds to the primary scintillation resulting from 22.1 and 25.0 keV X-rays absorbed in xenon (Fig.7). We believe that the peak obtained in the low energy region results from interactions in the presence of X-rays, such as luminescence and/or fluorescence of the detector materials as a result of X-ray and/or VUV photon interactions.

In order to extract the pulse-height distribution for primary scintillation, the low-energy peak was subtracted from the pulse-height distributions assuming a Gaussian shape (Fig. 8). Note that the fitted curves are very similar for 6 keV and 22 keV. The non-gaussian shape of the obtained primary scintillation distributions comes from solid angle effects as the amount of scintillation photons reaching the PMT depends on the position (depth) where the primary scintillation is produced.

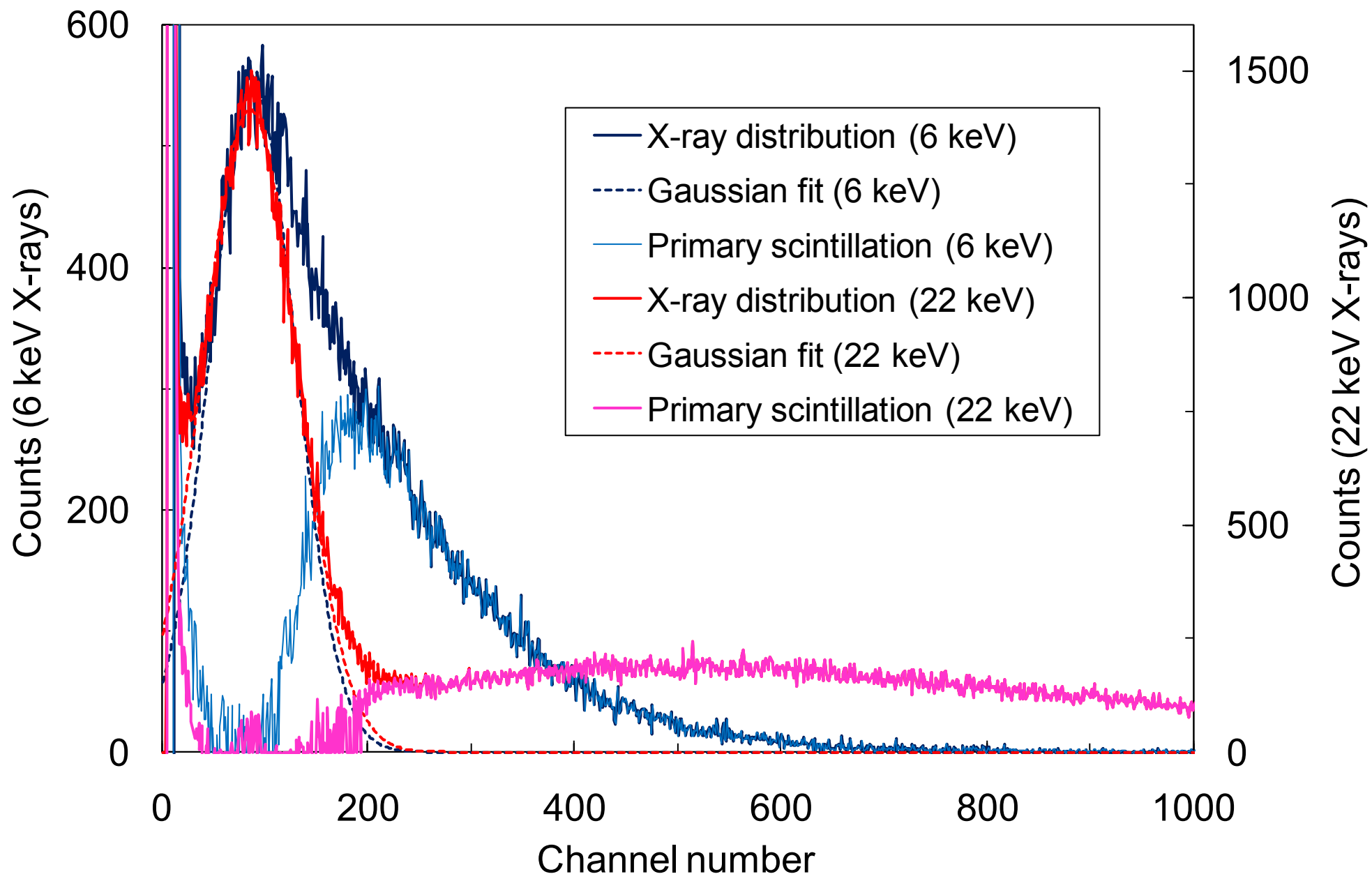

Fig. 8. Estimated pulse height distributions of primary scintillation produced by 5.9 keV and 22.1 keV X-rays absorbed in the xenon, obtained by subtraction of a Gaussian curve fitted to the low-energy region of the X-ray distribution.

The variation of the primary scintillation distributions obtained for 5.9 keV X-rays with the drift electric field and with the xenon pressure was investigated (Fig. 9). For these measurements, no electric field was applied to the scintillation region. The pulse-height distributions of Fig. 9 (a) were obtained at 1 bar. The distributions of Fig. 9 (b) were obtained at a drift electric field of 0.2 V $cm^{-1}$ $torr^{-1}$. As seen, no significant dependence was found on both the drift electric field and the pressure.

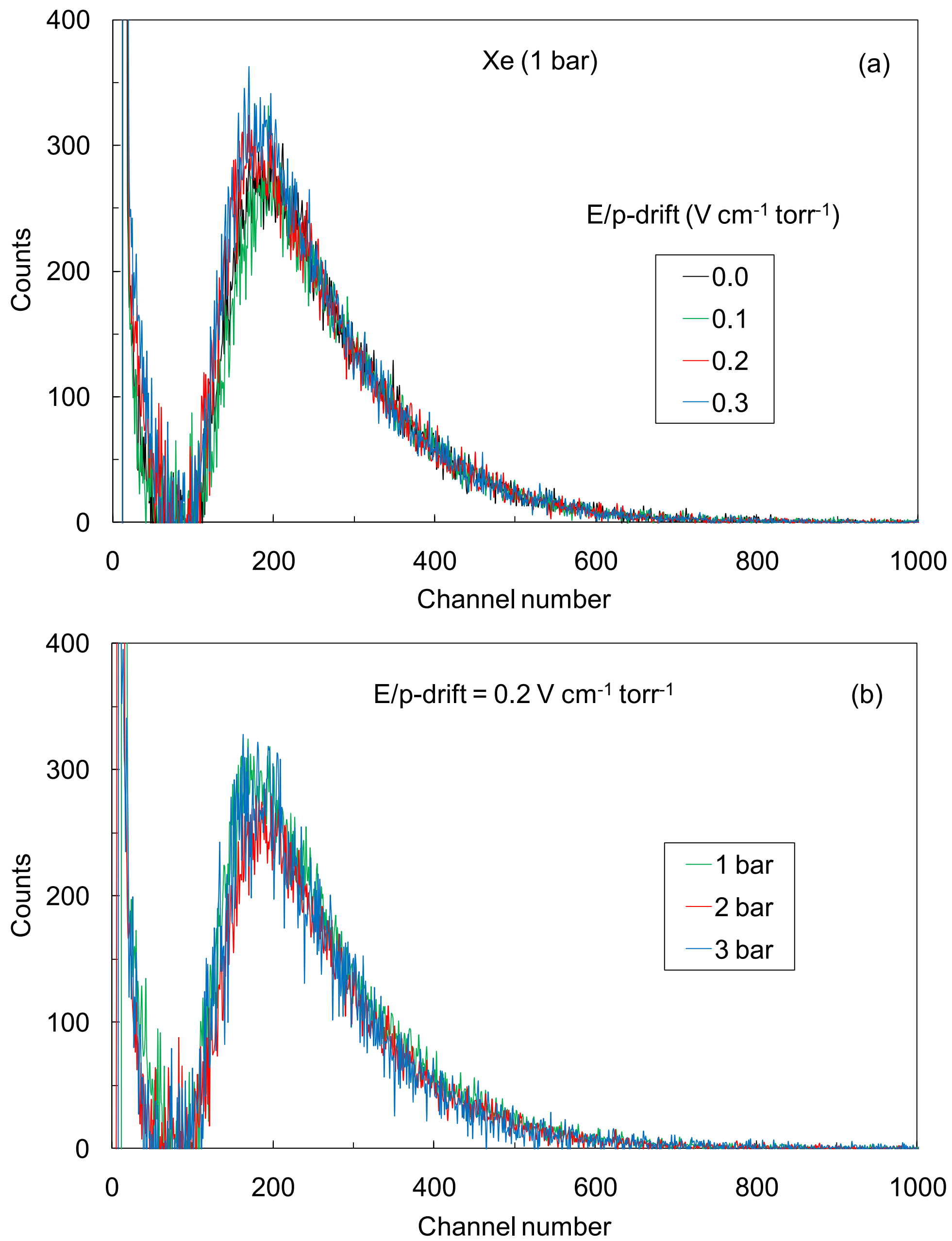

Fig. 9. Estimated primary scintillation distributions for 5.9 keV X-rays absorbed in the xenon: (a) for different drift electric fields; (b) for different gas pressures.

The results demonstrate promising operation characteristics of large volume xenon TPCs. No evidence of significant recombination at low drift electric field and high pressure was found.

## 5. Primary scintillation yield

The ratio between primary and secondary scintillation pulse amplitudes can be used to determine the absolute number of primary scintillation photons produced upon X-ray absorption, enabling the calculation of the average energy required to produce a primary scintillation photon ($W_s$). The primary and secondary scintillation pulse amplitudes are proportional to the average numbers of primary and secondary scintillation photons, respectively, reaching the PMT.

The number of secondary scintillation photons reaching the PMT is given by:

$$N_s = N_e\, T\, Y \frac{1}{4\pi} \int_D^{D+d} \Omega(x)\; dx \quad (3)$$

where $N_e$ is the average number of primary electrons produced per X-ray absorbed, $T$ is the grid optical transmission, $Y$ is the secondary scintillation yield per primary electron per cm, $\Omega(x)$ is the solid angle subtended by the PMT window at each scintillation point $x$ measured from the detector window, $D$ and $d$ are the thicknesses of the drift and scintillation regions, respectively.

Assuming a PMT photocathode with a circular shape of radius $R$, the solid angle $\Omega(x)$ along the drift and scintillation regions is given by:

$$\Omega(x) = 2\pi \frac{1-(D+d-x)}{\sqrt{(D+d-x)^2+R^2}}. \quad (4)$$

As the primary scintillation has to cross two grids, the number of primary scintillation photons reaching the PMT is given by:

$$N_p = N_0\, T^2\, \frac{\bar{\Omega}}{4\pi}, \quad (5)$$

where $N_0$ is the total number of primary scintillation photons produced per X-ray absorption, here called primary scintillation yield, and $\bar{o}$ is the average solid angle for primary scintillation. This average solid angle has to be weighted by the X-ray absorption rate $g(x)$, taken along the drift region, which is related to the X-ray absorption length $\lambda$:

$$g(x) = \frac{1}{\lambda} e^{-x/\lambda} \quad (6)$$

The average solid angle for primary scintillation is then given by:

$$\bar{\Omega} = \frac{\int_0^D g(x)\,\Omega(x)\, dx}{\int_0^D g(x)\, dx}. \quad (7)$$

From Eq. (3) and Eq. (5), we obtain:

$$\frac{N_s}{N_p} = \frac{N_e\, Y \int_D^{D+d} \Omega(x)\; dx}{N_0\, T\, \bar{\Omega}}. \quad (8)$$

This ratio is then equal to the ratio between the secondary and primary scintillation pulse amplitudes. $N_0$ can be determined from Eq. (8) since all other parameters in the second term can be calculated. The average energy required to produce a primary scintillation photon can be then estimated as $W_s = E_x/N_0$, $E_x$ being the energy of the incident X-ray.

The number of primary electrons produced by 5.9 keV X-rays is $N_e = E_x/w = 263$ assuming a $w$-value of 22.4 eV for xenon [15]. The secondary scintillation yield $Y$ is related to the electric field $E$ in the scintillation region and can be obtained from the equation [9]:

$$\frac{Y}{p} = 105\frac{E}{p} - 116 \quad (9)$$

where $Y/p$ is expressed in photons per electron per cm per bar and $E/p$ in V cm$^{-1}$ torr$^{-1}$. For $E/p$ = 2.0 V cm$^{-1}$ torr$^{-1}$, Eq. (9) gives $Y/p$ = 94 photons per electron per cm per bar.

The other parameters of Eq. (8) are $T$ = 0.84, $D$ = 3 cm, $d$ = 0.5 cm. The solid angle was calculated numerically assuming a PMT with a circular shaped photocathode of the same area ($R$ = 1.24 cm), and taking into account the absorption length of 5.9 keV X-rays in xenon (0.26 cm at 1 bar [16]). The solid angle parameters obtained for 1 bar are then:

$$\int_{D}^{D+d} \Omega(x)\ dx = 2.533 \text{ sr cm}; \quad \overline{\Omega} = 0.424 \text{ sr.}$$

The ratio between the secondary and primary scintillation pulse amplitudes is shown in Fig. 10 as a function of the drift electric field. These results were obtained from oscilloscope measurements, like in Fig. 5. In order to observe secondary scintillation, an electric field of 2.0 V cm$^{-1}$ torr$^{-1}$ was used in the scintillation region. Within the experimental errors, the ratio is approximately constant for drift electric fields between 0.2 and 0.8 V cm$^{-1}$ torr$^{-1}$. An average value within this interval was used for calculations.

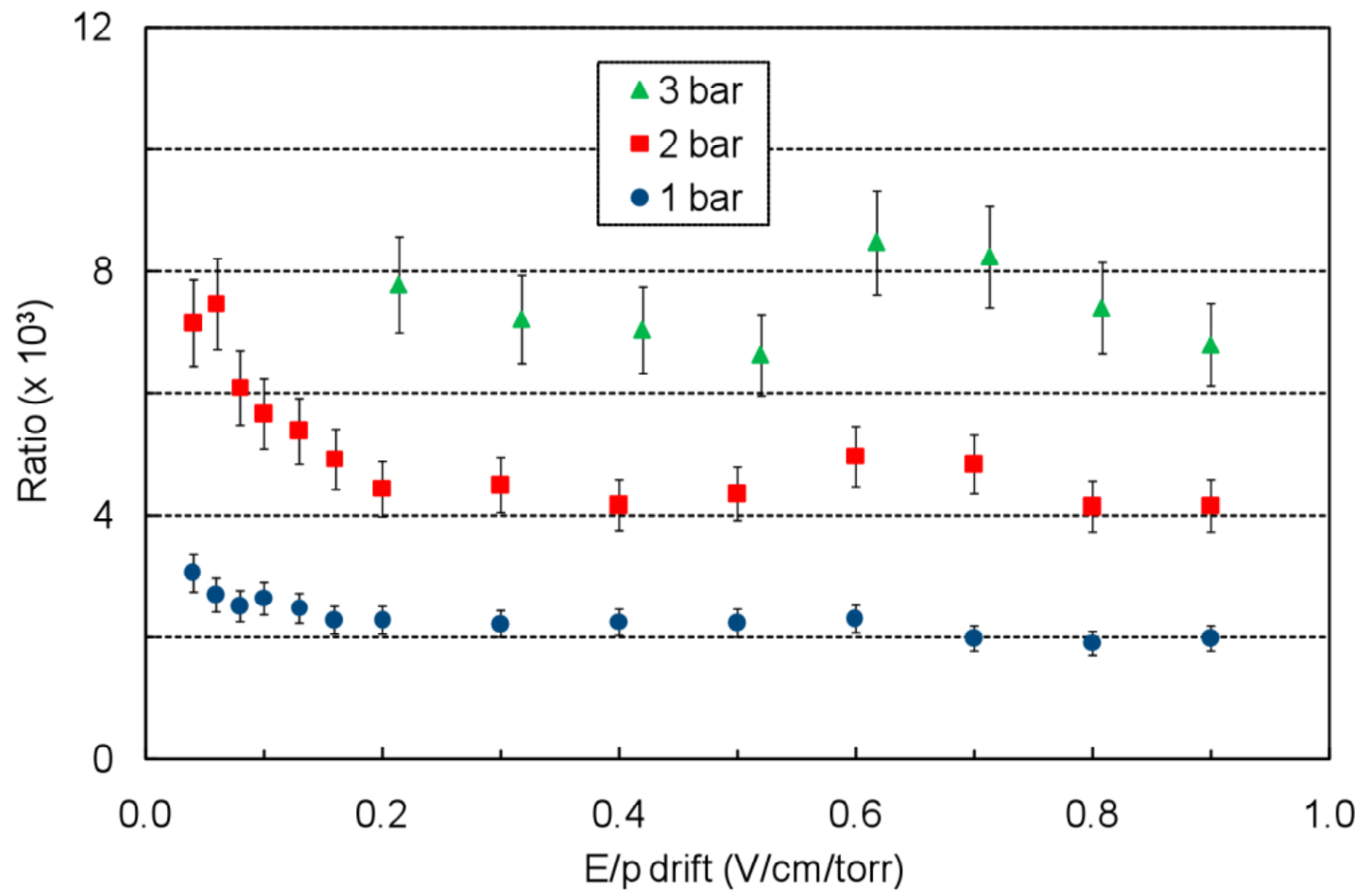

Fig. 10. Ratio between the secondary and primary scintillation pulse amplitudes as a function of the drift electric field, obtained from oscilloscope measurements, for 5.9 keV X-ray interactions.

Table I shows several parameters used in Eq. (8) to determine $N_0$ and $W_s$ for different gas pressures. As expected, the results don't have significant variations with pressure. The three values obtained are very compatible taking into account the respective errors. Averaging the results for the three different gas pressures (1, 2 and 3 bar), a final value $N_0$ = 81 ± 7 photons is obtained. Accordingly, the mean energy required to produce a primary scintillation photon is $W_s$ = 72 ± 6 eV. This value is lower than the previous measurement, 111 ± 16 eV [13], but is however similar to the value measured for 60 keV γ-rays at 20 bar, 76 ± 12 eV [17].

Table I – Parameters of Eq. (8) used to determine the primary scintillation yield in xenon for different gas pressures.

| Pressure (bar) | $N_s/N_p$ ($\times 10^3$) [Fig. 10] | Y (photons/$e^-$/cm) Ref. [9] | λ (cm) | $\overline{\Omega}$ | $N_0$ (photons) | $W_s$ (eV) |
|---|---|---|---|---|---|---|
| 1 | 2.2 ± 0.2 | 94 (±10%) | 0.261 | 0.424 | 80 ± 11 | 74 ± 10 |
| 2 | 4.5 ± 0.4 | 188 (±10%) | 0.131 | 0.389 | 85 ± 11 | 69 ± 9 |
| 3 | 7.5 ± 0.9 | 282 (±10%) | 0.087 | 0.379 | 79 ± 12 | 75 ± 12 |

In the present conditions, about 2 primary scintillation photons were detected in the PMT per 5.9 keV X-ray absorbed in xenon. For NEXT, double beta decay events have 2458 keV energy. The total number of primary scintillation photons produced in xenon will be about $2458\times10^3/72 = 3.4\times10^4$. As the PMT dimensions are much smaller than the distance to the interaction point, the solid angle subtended by the PMT is approximately given by $\Omega = A/d^2$ , where $A$ is the PMT active area and $d$ is the distance to the interaction point. For $d$ = 50 cm, the fractional solid angle is then $\Omega/4\pi = (2.2/50)^2/4\pi = 1.5\times10^{-4}$. This means that about 5 photons should reach the PMT. Furthermore, a large number of PMTs will be used and the sum of all PMT signals will make the signal large enough to be detected above the noise level. In addition, the visible light and X-ray background observed in our detector will not be present in the NEXT detector.

The number of photoelectrons produced in the PMT per primary electron crossing the scintillation region ($L$) can be estimated as it is determined by the secondary scintillation yield, Eq. (9), the average solid angle subtended by the PMT for secondary scintillation, the grid transmission and the quantum efficiency of the PMT (30% [5]). A value of $L$ = 20 is obtained for E/p = 5 V cm$^{-1}$ torr$^{-1}$ in the scintillation region, which is in good agreement with that obtained from the energy resolution values (section 3).

## 6. CONCLUSIONS

The performance of a Hamamatsu R8520-06SEL photomultiplier, used as VUV photosensor in a xenon GPSC, has been investigated, demonstrating that this PMT is a good candidate for the scintillation readout in the TPC to be used in NEXT.

The PMT high performance for secondary scintillation detection was demonstrated. An energy resolution of 8.0% (FWHM) was obtained for 5.9 keV X-rays absorbed in the xenon, similar to GPSCs instrumented with larger PMTs, demonstrating the very low statistical variance of electroluminescent gain.

Primary scintillation measurements have been carried out. The pulse-height distributions obtained in the MCA don't have significant variations with the electric field in the drift region and with the xenon pressure, demonstrating negligible recombination luminescence. Amplitude measurements of the primary scintillation produced by 5.9 keV X-rays were possible using an averaging process in the oscilloscope and triggering at the corresponding secondary scintillation pulse. These oscilloscope measurements allowed a determination of the primary scintillation yield in xenon gas. An average of 81 ± 7 primary scintillation photons produced by 5.9 keV X-rays absorbed in xenon was obtained. The average energy required to produce a primary scintillation photon in xenon was deduced, resulting $W_s = 72 \pm 6$ eV. This value is lower than the previous measurement, 111 ± 16 eV [13], but is however similar to that measured for 60 keV γ-rays at 20 bar, 76 ± 12 eV [17]. Since the production of primary scintillation results from collisional processes of the photoelectrons and other Auger and shake-of electrons with the gas atoms, it is expected that *Ws* does not vary significantly with the gas pressure.

Measurements were carried out only up to 3 bar pressures due to a limitation of our experimental setup. A new chamber was built at IFIC (Valencia) in order to study the PMT response at higher pressures. The aim is to verify the PMT performance at the TPC pressure of 10 bar.

## ACKNOWLEDGMENTS

This work was supported by FCT (Portugal) and FEDER through project PTDC/FIS/103860/2008. E.D.C. Freitas acknowledges grant SFRH/BD/46711/2008 from FCT. C.M.B. Monteiro acknowledges grant SFRH/BD/25569/2005 from FCT. M. Ball, J.J. Gómez-Cadenas and N. Yahlali acknowledge the Spanish MICINN for the Consolider-Ingenio grants CSD2008-00037 and CSD2007-00042 and the research grants FPA2009-13697-C04-04 and FPA2009-13697-C04-B23/12. D.R. Nygren acknowledges support by the Director, Office of Science, Office of High Energy Physics, of the U.S. Department of Energy under contract DE-AC02-05CH11231.